\documentclass[trackchanges]{aastex7}
\usepackage{amsmath}	% Advanced maths commands
\usepackage{multirow}

\begin{document}

%\title{To Disentangling Disk and Jet Contribution in Blazars from Optical Spectroscopic Data: Applications to PKS 1510-089 and PKS 0736+017}

%\title{Decomposition of Disk and Jet Emission in Blazars Based on Optical Spectral Shape}
\title{A novel reverberation mapping method for  blazars}

\correspondingauthor{Yunguo Jiang (jiangyg@sdu.edu.cn)}
\author[0009-0007-1984-6603]{Junhao Deng}
\affiliation{Shandong Provincial Key Laboratory of Optical Astronomy and Solar-Terrestrial Environment, Institute of Space Sciences, Shandong University, Weihai, 264209, People’s Republic of China}
\affiliation{School of Space Science and Technology, Shandong University, Weihai 264209, People’s Republic of China}
\email{junhaodeng@foxmail.com}

\author{Lizhi Liu}
\affiliation{Shandong Provincial Key Laboratory of Optical Astronomy and Solar-Terrestrial Environment, Institute of Space Sciences, Shandong University, Weihai, 264209, People’s Republic of China}
\affiliation{School of Space Science and Technology, Shandong University, Weihai 264209, People’s Republic of China}
\email{380570843@qq.com}
\author[0000-0001-7142-7667]{Yifan Wang}
\affiliation{College of Physics and Electronic Information, Dezhou University, Dezhou 253023, People’s Republic of China}
\email{wangyf@dzu.edu.cn}

\author[0000-0003-2679-0445]{Yunguo Jiang}
\affiliation{Shandong Provincial Key Laboratory of Optical Astronomy and Solar-Terrestrial Environment, Institute of Space Sciences, Shandong University, Weihai, 264209, People’s Republic of China}
\affiliation{School of Space Science and Technology, Shandong University, Weihai 264209, People’s Republic of China}
\email{jiangyg@sdu.edu.cn}

%\collaboration{all}{The Terra Mater collaboration}

%% Use the \collaboration command to identify collaborations. This command
%% takes an optional argument that is either a number or the word "all"
%% which tells the compiler how many of the authors above the command to
%% show. For example "\collaboration[all]{(DELVE Collaboration)}" wil include
%% all the authors above this command.
%%
%% Mark off the abstract in the ``abstract'' environment. 
\begin{abstract}
Reverberation mapping (RM) is the most promising method to measure the masses of supermassive black holes in the center of active galaxy nuclei (AGNs). However, the dominant jet component hinders the application of RM method for blazars. In this work, we present a new algorithm to disentangle the contribution of the accretion disk from that of the relativistic jet in blazars by analyzing the spectral break of the optical spectroscopic data. 
We applied this method to two flat-spectrum radio quasars (FSRQs), PKS 1510-089 and PKS 0736+017. In PKS 1510-089, the variability of the H$\gamma$ line is delayed with respect to the disk emission by approximately 94 days, while the H$\beta$ line shows a lag of about 111 days relative to the disk. In PKS 0736+017, the H$\gamma$ variability is delayed with respect to the disk by roughly 66 days, and the H$\beta$ line exhibits a lag of about 67 days. Based on these measured time lags, we estimate black hole masses of $\sim1.4\times10^{8}\,M_{\odot}$ for PKS 1510-089 and $\sim8.1\times10^{7}\,M_{\odot}$ for PKS 0736+017.
This method paves the way to apply the RM method for blazars, and improves the understanding of disk and jet activities.

%using data from the Steward Observatory, we identify a spectral break arising from the superposition of jet and disk emission. 
%Using a model combining a power-law jet spectrum with a Shakura–Sunyaev disk to reproduce the break, we derive separate light curves for the two components. 

%%Our results show that the emission line variations lag those of the disk component, confirming the physical consistency of the decomposition. Compared with the commonly used non-thermal dominance (NTD) approach, our method preserves more data and provides a more robust way to trace the BLR response in jet-contaminated blazars.

\end{abstract}

%% Keywords should appear after the \end{abstract} command. 
%% The AAS Journals now uses Unified Astronomy Thesaurus (UAT) concepts:
%% https://astrothesaurus.org
%% You will be asked to selected these concepts during the submission process
%% but this old "keyword" functionality is maintained in case authors want
%% to include these concepts in their preprints.
%%
%% You can use the \uat command to link your UAT concepts back its 
\keywords{\uat{Active galactic nuclei}{16}; \uat{Blazars}{164}; \uat{Supermassive black holes}{1663}}

%% From the front matter, we move on to the body of the paper.
%% Sections are demarcated by \section and \subsection, respectively.
%% Observe the use of the LaTeX \label
%% command after the \subsection to give a symbolic KEY to the
%% subsection for cross-referencing in a \ref command.
%% You can use LaTeX's \ref and \label commands to keep track of
%% cross-references to sections, equations, tables, and figures.
%% That way, if you change the order of any elements, LaTeX will
%% automatically renumber them.

\section{Introduction} 

Active galactic nuclei (AGNs) are among the brightest objects in the universe, and have a supermassive black hole (SMBH) at their centers. The accretion process of SMBH leads to distinct radiative structural components, including the accretion disk and the broad-line region (BLR). The BLR consists of high-velocity gas, which is mainly photoionized by the ultraviolet (UV) and optical photons from the accretion disk. The profile of the broad emission lines traces the kinematic and spatial distribution of the gas \citep{1995PASP..107..803U}. The luminosity variation of the accretion disk will lead to the time-delay responses of the emission-lines (e.g., \citealt{1992ApJ...392..470P,2002ApJ...581..197P,2004ApJ...606..749K,2012ApJ...747...30P,2015ApJS..217...26B,2015ApJ...804..138H}). This time delay can be measured using methods such as the discrete correlation function (DCF; \citealt{1988ApJ...333..646E}), the interpolated cross-correlation function (ICCF; \citealt{1998ApJ...501...82P}), the local cross-correlation function (LCCF; \citealt{1999PASP..111.1347W}), and the running optimal average (ROA) method \citep{2021MNRAS.508.5449D}, among others, to estimate the size of the BLR. Assuming the virial motion of BLR gas, the mass of the SMBH can be estimated from the lag and the full width at half maximum (FWHM) or line dispersion ($\sigma_{\rm line}$) of the broad emission lines, with the inclusion of a virial factor $f$ that accounts for the geometry and kinematics of the BLR. This technique is known as reverberation mapping (RM, \citealt{1993PASP..105..247P})

%Assuming the virial motion of BLR gas, the mass of SMBH can be derived via the lag and the full width at half maximum (FWHM) of the broad emission lines, known as the reverberation mapping (RM, \citealt{1993PASP..105..247P}). 
 
%However, it is difficult to apply the RM process to blazars, a special class of AGNs with their relativistic jet directed almost along the observer’s line of sight. The jet emission normally is Doppler boosted, and highly variable. In the optical–UV band, the emission is composed of radiation from both the accretion disk and the jet \citep{2010MNRAS.402..497G}. Therefore, disentangling the varied disk and jet components is a key to measure the SMBH mass in blazars  by using the RM process.

However, applying the RM technique to radio-loud AGNs, particularly blazars, remains challenging. For blazars, the jet emission is dominant and highly variable, making it difficult to disentangle variation of the accretion disk from that of jet. Consequently, over the past few decades, RM studies have been successfully carried out for only a limited number of blazars. For the low-synchrotron-peaked source, such as 3C 273, the jet emission is week in the optical band \citep{2021JHEAp..29...31P}, and the optical emission is dominated by the disk, making RM feasible. The central black hole mass of 3C 273 is estimated to be approximately $4.10\times10^{8}\ M_\odot$ \citep{2019ApJ...876...49Z}.
On the other hand, for other sources, the non-thermal dominance (NTD) parameter is employed to classify the data into jet-dominated and disk-dominated states \citep{2012ApJ...748...49S}. This method assumes that when $\mathrm{NTD} < 2$, the specific fluxes at $5100\,\text{\AA}$ is dominated by the accretion disk. When performing the lag correlation analysis, data with $\mathrm{NTD} > 2$ are excluded. \cite{2024ApJ...977..178A} and \cite{2025ApJ...979..227A} used the NTD method to estimate the central black hole mass of PKS 1510-089 to be $2.85\times10^{8}\ M_\odot$. This method relies on a relatively straightforward threshold-based criterion, which may limit its robustness. In other words, even when the NTD is below 2, variability from the jet may still contribute to the observed fluxes, potentially disturbing the correlation analysis. 
Moreover, for sources where the NTD method cannot be applied, that is, when the majority of data have $\mathrm{NTD} > 2$, the only option is to directly measure the time lag between the $V$-band light curve and the broad emission lines. For example, \cite{2022MNRAS.516.2671P} measured a time lag of about 70 days between the H$\beta$ line and the $V$-band light curve in PKS 0736+017, and used this lag to report a central black hole mass of approximately $7.32\times10^{7}\ M_\odot$. In their analysis, however, the correlation coefficient of the ICCF was relatively low, with a peak value below 0.5.

%The Non-Thermal Dominance (NTD) parameter was used to classify the data into jet-dominated and disk-dominated states. They assumed that when $\mathrm{NTD} < 2$, the specific fluxes at $5100\,\text{\AA}$ is dominated by the accretion disk. This method relies on a relatively straightforward threshold-based criterion, which may limit its robustness. In other words, even when the NTD is below 2, variability from the jet may still contribute to the observed fluxes, potentially disturbing the correlation analysis.} %Besides, the NTD less than 2 cases are unpredictable and mostly happen during quiescent states. These ingredients significantly intervene the application of the RM procedure for blazars.    

In this paper, we present a approach to distinguish the contributions of the accretion disk and the jet, and pave a way for the application of RM for a subclass of blazars. Blazars with broad emission lines are classified as flat-spectrum radio quasars (FSRQs). According to the broad band spectral energy distribution (SED), FSRQs are typically categorized as the low-synchrotron-peaked blazars, with the peak frequency of the synchrotron bump lying below $10^{14}$ Hz \citep{2010ApJ...715..429A}. Thus, in the optical observation window, the jet emission exhibits a power-law spectrum with negative spectral index in the $\log \nu F_{\nu}$ versus $\log \nu$ plot. The accretion disk, contributing the big blue bump (BBB) in the SED, usually peaks in the UV band and shows a positive spectral index in the optical window. Therefore, we can predict that the spectroscopic data is expected to show a spectral break when the disk and jet emissions differ by less than a magnitude. Thus, by quantitatively modeling the spectral break, we are able to disentangle the contributions from the accretion disk and the jet. In this work, we select PKS 1510-089 and PKS 0736+017 as case studies to explore the feasibility of this new approach, since both targets have the abundant publicly available spectroscopic data and evident spectral break. The structure of this paper is organized as follows. Section \ref{sec:2} details the data analysis and processing. Section \ref{sec:3} presents the results, and Section \ref{sec:4} offers discussion and conclusions. In addition, the redshift values used in this study are $z = 0.361$ for PKS 1510-089 \citep{1966ApJ...145..654B} and $z = 0.189$ for PKS 0736+017 \citep{1967ApJ...147..837L}.

\section{Data Reduction} \label{sec:2}
All spectroscopic data used in this study are from the Steward \text{Observatory}\footnote{\url{https://james.as.arizona.edu/~psmith/Fermi/}}. These data were obtained with the 2.3 m Bok Telescope on Kitt Peak and the 1.54 m Kuiper Telescope on Mount Bigelow in Arizona, using the SPOL spectropolarimeter \citep{1992ApJ...398L..57S}. A 600 $\rm mm^{-1}$ grating provided a spectral range of 4000-7550 \AA\ with a dispersion of 4 \AA\ /pixel. Depending on the slit width, the spectral resolution was typically 16–24 \AA\ . Flux calibration followed the procedure described in \cite{2009arXiv0912.3621S}. The spectra were first calibrated using sensitivity functions derived from spectrophotometric standard stars, and then re-scaled to match the synthetic V-band photometry for each night. The dataset employed in this work has previously been reported in earlier studies (\citealt{2022MNRAS.516.2671P,2024ApJ...977..178A,2025ApJ...979..227A}). From the public website of the Steward Observatory, we downloaded 371 spectra of PKS 1510-089 (observed between MJD 54,830 and 58,307), 130 spectra of PKS 0736+017 (observed between MJD 56,989 and 58,251). The Galactic interstellar extinction and reddening of all spectra were corrected by the procedure given in \cite{1989ApJ...345..245C}, and $E(B-V)$ of sources refers to the NASA/IPAC Extragalactic \text{Database}\footnote{\url{https://ned.ipac.caltech.edu/}}. The spectra spans a frequency range from approximately $10^{14.60}$ to $10^{14.88}$ Hz.

\subsection{Spectral Break} \label{sec:2.1}

The top-left and top-right panels of Figure 1 display spectra of PKS 0736+017 and PKS 1510–089 at three different flux levels. At the low-flux states, the spectra exhibit a concave shape in the plane of of $\log \nu$ versus $ \log \nu F_\nu$, namely a spectral break. This spectral break shifts, becomes less prominent and gradually disappears at the high-flux levels. The detailed analysis of the spectral break and its dependence on the flux level will be presented in Section 2.4.

To illustrate the origin of the spectral break, we show the low-flux state broadband SEDs of PKS 1510–089 and PKS 0736+017, compiled from historical archival data downloaded from the Space Science Data \text{Center}\footnote{\url{https://tools.ssdc.asi.it/SED/}} (SSDC) in bottom panels of Figure 1. 
We reproduced the broadband SEDs using a one-zone leptonic model. Details of the model and its parameters are provided in Appendix A. Within the observing band, the total energy flux consists of two components, a red component and a blue component. The red component corresponds to the non-thermal synchrotron emission from the jet, while the blue component arises from thermal emission of the accretion disk.

In general, the variability of the jet is more significant than that of the accretion disk. When the target changes from a quiescent state to an active state, both the peak flux and frequency of the jet component increase \citep{2024ApJ...966...65W,2025MNRAS.536.1251W,2025ApJ...983..128D}.  Consequently, the spectral break evident in low-flux states disappears in the high-flux states. 

\begin{figure*}
	\centering
         \includegraphics[width=0.49\linewidth]{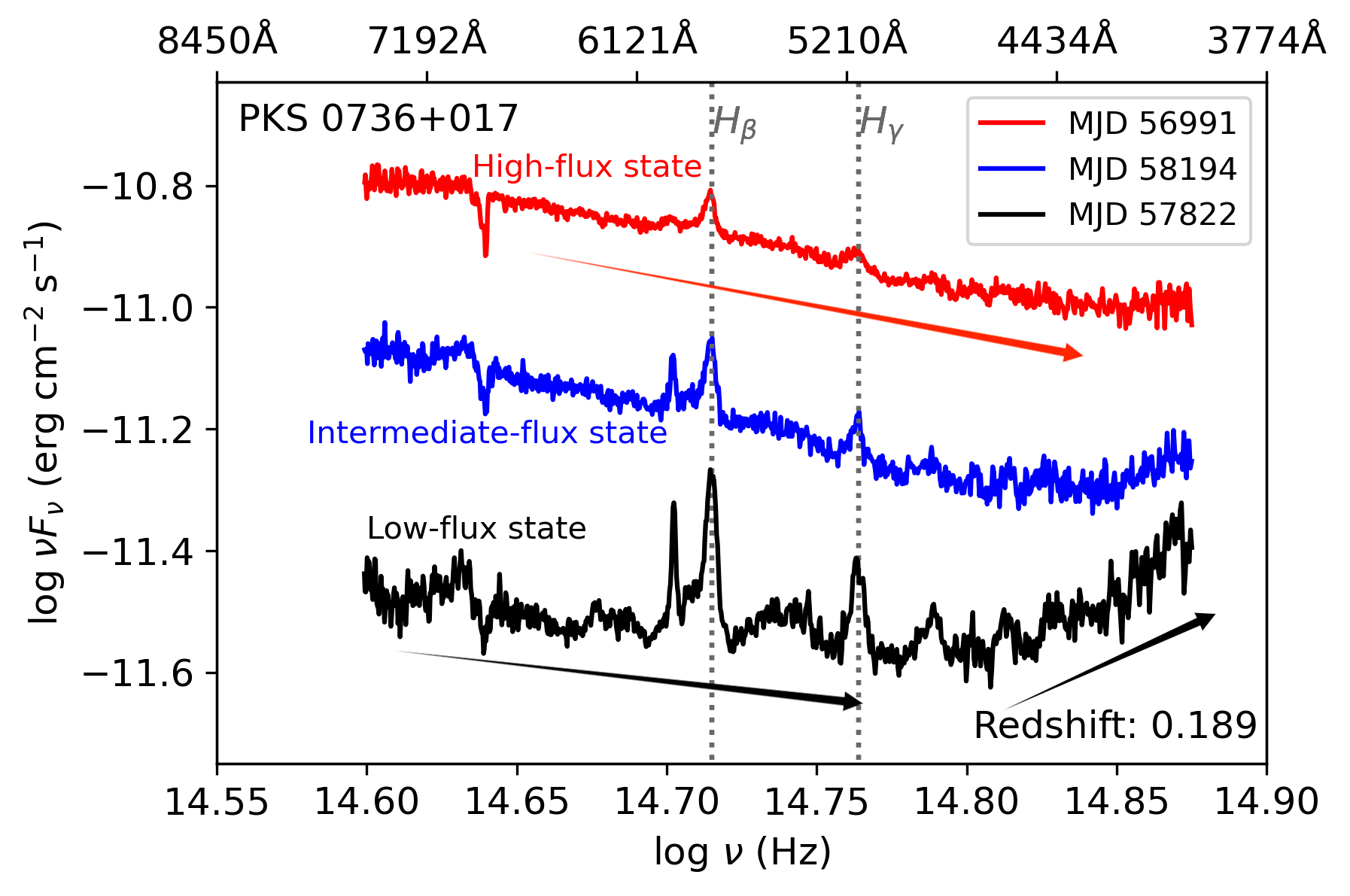}
         \includegraphics[width=0.49\linewidth]{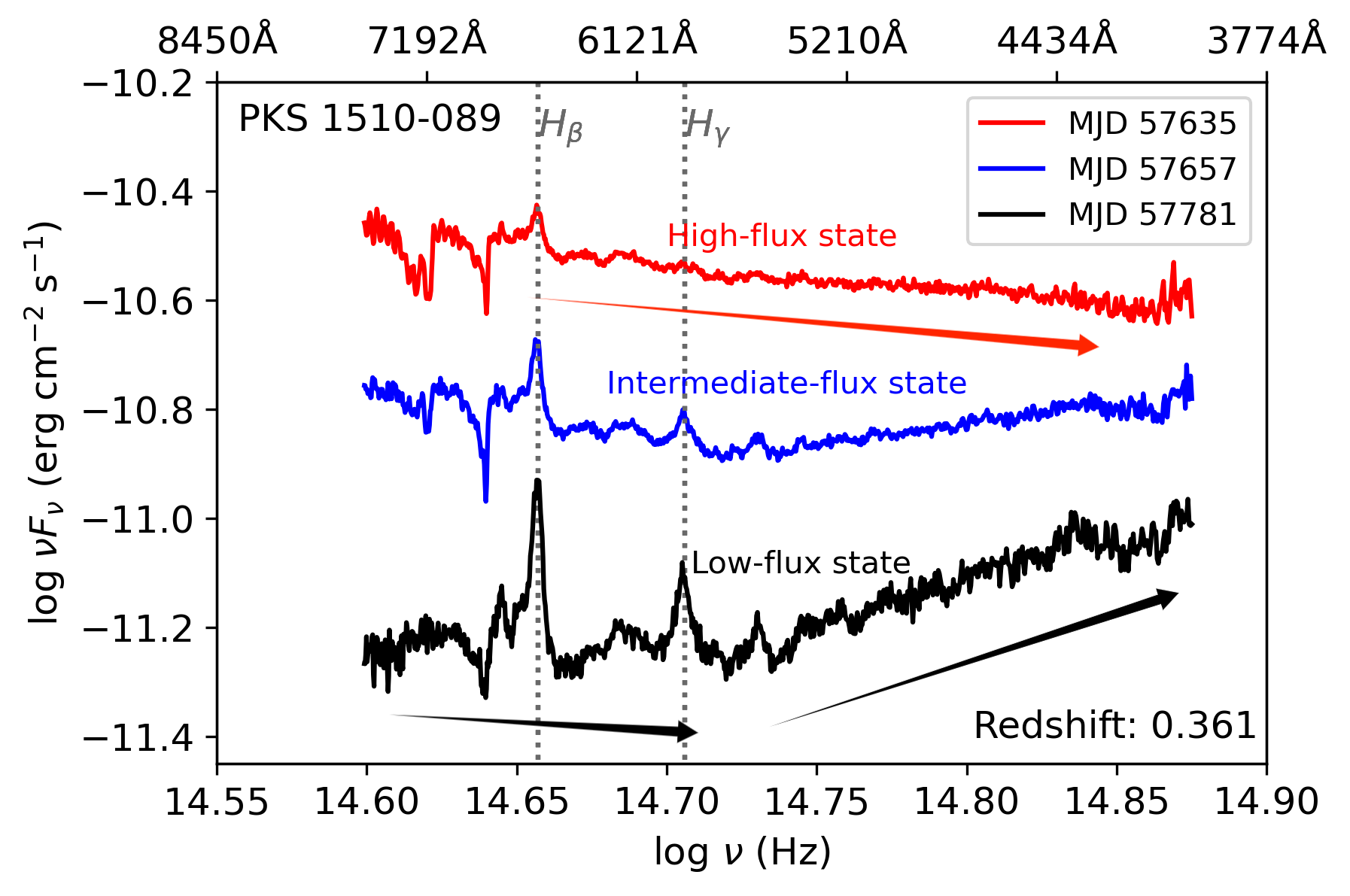}
         \includegraphics[width=0.49\linewidth]{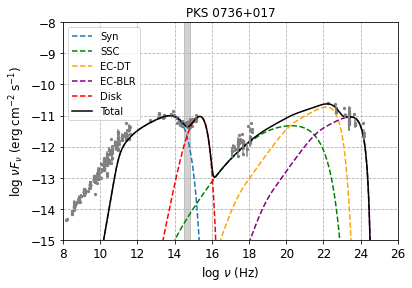}
         \includegraphics[width=0.49\linewidth]{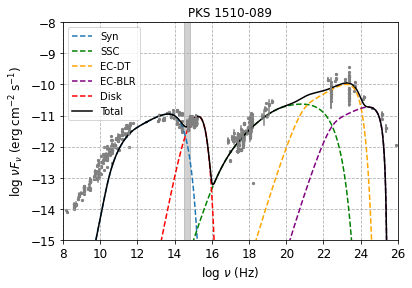}
    \caption{Top panels: Representative optical spectra of PKS 0736+017 (left) and PKS 1510-089 (right) in three different flux states. The arrows approximately mark the local spectral slope, serving as visual guides. Bottom panels: Broadband SEDs in the low-flux state for PKS 0736+017 (left) and PKS 1510-089 (right), compiled from archival data and modeled by using the one-zone leptonic model (see Appendix A). The reduced $\chi^2$ values are 114.03 for PKS 0736+017 and 153.06 for PKS 1510-089, with radio data excluded from the calculation. The shaded gray regions mark the optical observation window. The overlap of jet and disk components confirms that both contribute to the observed optical emission. }
    \label{fig:guangpuduoshiqi}
\end{figure*}

\subsection{Emission Lines Fitting Scheme} \label{sec:2.2}

In this section, we describe the method used to obtain the light curve of emission lines through the spectral fitting procedure. First of all, the observed spectroscopic data were shifted to the rest-frame wavelengths based on their redshifts.

\begin{figure*}
	\centering
         \includegraphics[width=0.9\linewidth]{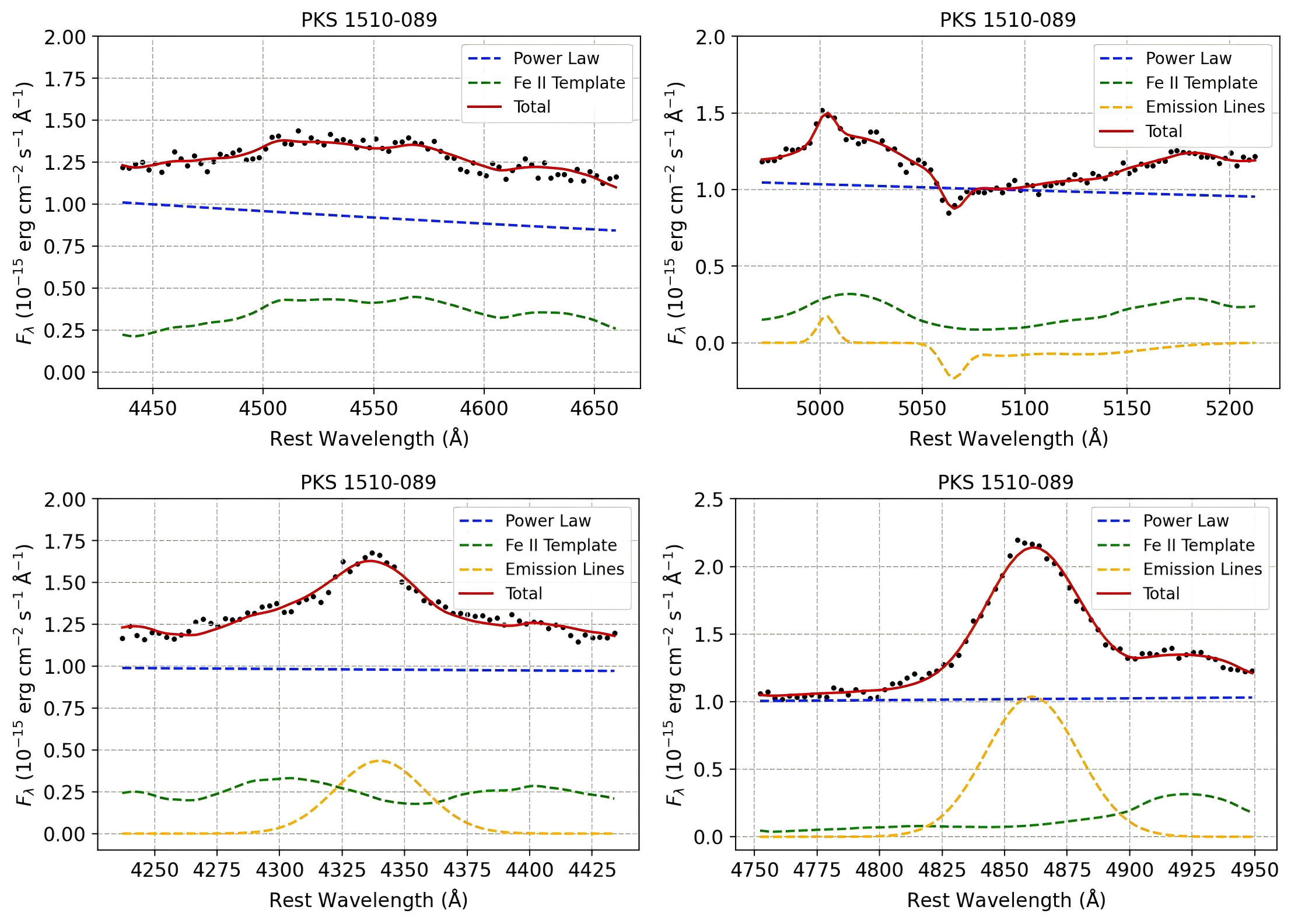}
    \caption{An example of the spectral decomposition for PKS 1510–089, using the spectrum observed on MJD 58253.
    %Multi-component spectral fitting of the average optical spectra for PKS 0736+017 (left) and PKS 1510-089 (right). The model consists of a power-law continuum (representing the combined jet and disk emission), an Fe\,II template, a host galaxy template, and multiple Gaussian profiles for the emission lines. For PKS 0736+017, the fit includes the H$\beta$ $\lambda4861$, [O\,III] $\lambda4959$, and [O\,III] $\lambda5007$ emission lines within the rest-frame wavelength range of 4800--5050\,\AA. For PKS 1510-089, due to strong telluric absorption near H$\beta$, the fitting is performed in the 4250--4500\,\AA\ range, including H$\gamma$ $\lambda4340$, [O\,III] $\lambda4363$, and He\,I $\lambda4471$ lines. 
  } 
    \label{fig:fashexianguangpulizi}
\end{figure*}

For both targets, the H$\beta$ and H$\gamma$ broad lines are present within the optical window, as seen in top panels of Figure \ref{fig:guangpuduoshiqi}. In this study, we extracted the light curves of these two lines. Since fitting the entire optical range spectrum with a single power-law function tends to either underestimate the high-frequency end when the low-frequency part is well fitted, or vice versa,  we adopted a segmented fitting strategy for each observation ID. Specifically, we divided the spectrum into four wavelength regions: (i) 4435–4685 \AA,  dominated by Fe II features, used to determine the parameters of the Fe II emission lines with the template taken from \citet{1992ApJS...80..109B}; (ii) 4200–4435 \AA, with the Fe II template parameters from step (i), to measure the H$\gamma$ flux; (iii) 4750–4950 \AA, to measure the H$\beta$ flux; and (iv) 4980–5230 \AA, used to determine the 5100 \AA\ continuum flux, also including the Fe II contribution fixed from (i). The [O III] line was fixed during the fitting, with its parameters obtained from the average spectrum. For PKS 1510–089, a prominent telluric absorption feature is present near 5100 \AA. We corrected for this feature following the same procedure as \cite{2024ApJ...977..178A}, by modeling it with three Gaussian components. The fitting was performed in Python, employing the minimize function from SciPy \citep{2020NatMe..17..261V} to obtain the best-fit parameters. Figure \ref{fig:fashexianguangpulizi} illustrates an example of the spectral decomposition for PKS 1510–089, using the spectrum observed on MJD 58253. Through the numerical integration of the Gaussian functions, we obtained the flux of the emission lines. The light curves of the emission lines and 5100 \AA\; continuum are shown in Figure \ref{fig:lc}.  %Finally, we collect all these fluxes to obtain the light curves of emission lines.}
%鉴于将4000埃-5200埃的数据作为整体拟合，用一个幂律代表连续谱会造成拟合好了低频则低估高频端或者拟合好高频就低估低频,这是因为连续谱是一个spectral break的谱，因此，对每一个观测ID，我们都分了四段进行拟合，第一段先拟合4435-4680埃确定铁线的参数，这一段几乎都是Fe II的特征。然后我们固定了铁线的参数，拟合4200-4435埃，得到Hgamma的流量，再然后仍然固定铁线，我们拟合4750-4950埃，得到Hbeta的流量，最后我们拟合4980-5230埃，得到5100埃连续谱的流量，[O III]我们没有固定，但是实际拟合得到的通量几乎没有明显变化。 
%In correcting for the telluric lines, we followed the method of Amador-Portes et al.(2024, https://doi.org/10.3847/1538-4357/ad8ddd), and adopted three Gaussian functions to model the absorption features. The corresponding results are now included in the revised manuscript.

%For PKS 0736+017, we selected the H$\beta$ line for analysis and extracted spectra from 4800~\AA\ to 5050~\AA\ , since the H$\beta$  line is prominent than H$\gamma$. In the case of PKS 1510-089, there is strong telluric absorption features near 4861~\AA. Thus, we chose to fit the H$\gamma$ instead, using the spectra from 4250~\AA\ to 4500~\AA. The spectra at all epochs were stacked to produce the time-average spectra for each target, shown in Figure 2.

\subsection{Disentangle the disk and jet} \label{sec:2.3}

To disentangle the emission from the accretion disk and the jet, we employed a broader spectral range (log $\nu$[Hz] from 14.60 to 14.88) rather than the narrow range used in Section \ref{sec:2.2}. The jet component was modeled with a power-law spectrum with a negative spectral index, while the accretion disk was modeled by the Shakura-Sunyaev model \citep{1973A&A....24..337S}. All disk parameters except the luminosity were fixed based on the fitting result of the broad band SEDs shown in bottom panels of Figure \ref{fig:guangpuduoshiqi}. 

\begin{figure*}
	\centering
         \includegraphics[width=0.45\linewidth]{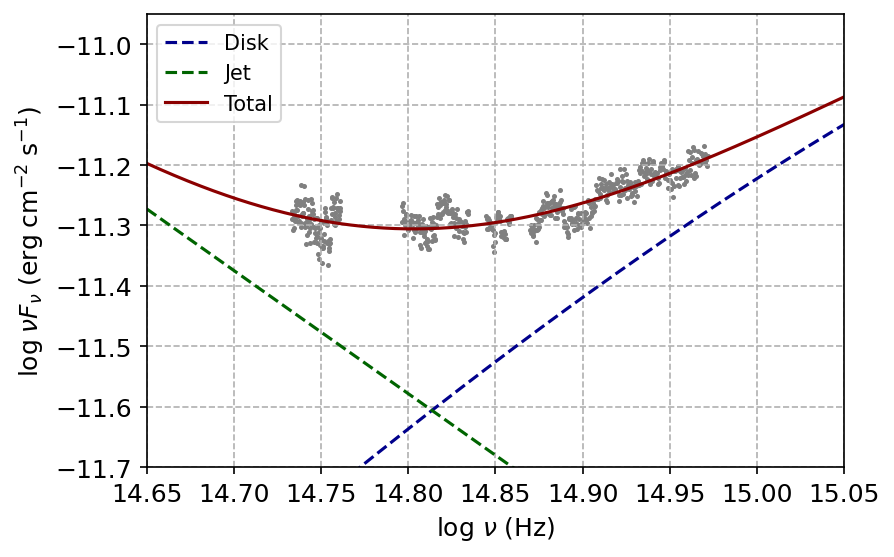}
    \caption{Spectral decomposition of PKS 1510-089 on MJD 58253, shown as a representative example to illustrate the disk–jet emission separation method. The gray points show the optical spectrum after removal of the Fe II emission, prominent emission lines, and atmospheric absorption features. The green dashed line represents the power-law component corresponding to non-thermal jet emission, while the blue dashed line represents thermal emission from the accretion disk, modeled with the Shakura-Sunyaev model. The red solid line shows the total emission combining both components. The reduced $\chi^{2}$ value of the fit is 0.28.} 
    \label{fig:fenjienihe}
\end{figure*}

We first subtracted the contributions of the Fe II emission with parameters adopted from the fit in Section \ref{sec:2.2}. Second, we excluded the spectral ranges containing prominent emission lines and atmospheric absorption. Specifically, we masked the 4800-4900~\AA\ around H$\beta$ $\lambda$4861, the 4950-5050~\AA\ around [O III] $\lambda$5007, and the 4300-4400~\AA\ around H$\gamma$ $\lambda$4340. The spectra at different epochs may have different ratio of jet over disk components. In particular, during high-flux states, as shown in the top panels of Figure \ref{fig:guangpuduoshiqi}, the disk contribution becomes negligible due to the disappearance of the spectral break, and including a disk component in the fit does not yield reliable results. Thus, we consider three fitting models, i.e., (1) only the jet, (2) only the disk, and (3) both the jet and the disk.  %The fitting was performed using MCMC methods.
For each case, we calculated the value of Akaike Information Criterion (AIC), which considers the trade-off between goodness of fit and model complexity. The lower value of AIC indicates a better result. 
In many cases, the case (3) gives the lowest AIC, suggesting that the jet and disk have comparable contributions, as shown in Figure \ref{fig:fenjienihe}. At high flux states, case (1) gives the lowest AIC, which indicates the dominance of jet component. However, across all observed spectra of PKS 1510-089 and PKS 0736+017, we did not find any epoch in which the case (2) yielded the lowest AIC. In addition, the jet-only model (case 1) provides the lowest AIC in 28\% of the observations for PKS 1510-089, and in 22\% of the observations for PKS 0736+017. Figure \ref{fig:fenjienihe} presents an example of the fit by case 2. Finally, the fluxes of the disk and jet components within the observed frequency range were calculated via numerical integration. The light curves of the disk and jet component for the two targets are plotted in Figure \ref{fig:lc}.

\subsection{NTD parameters} \label{sec:2.4}

\begin{figure*}
	\centering
         \includegraphics[width=0.99\linewidth]{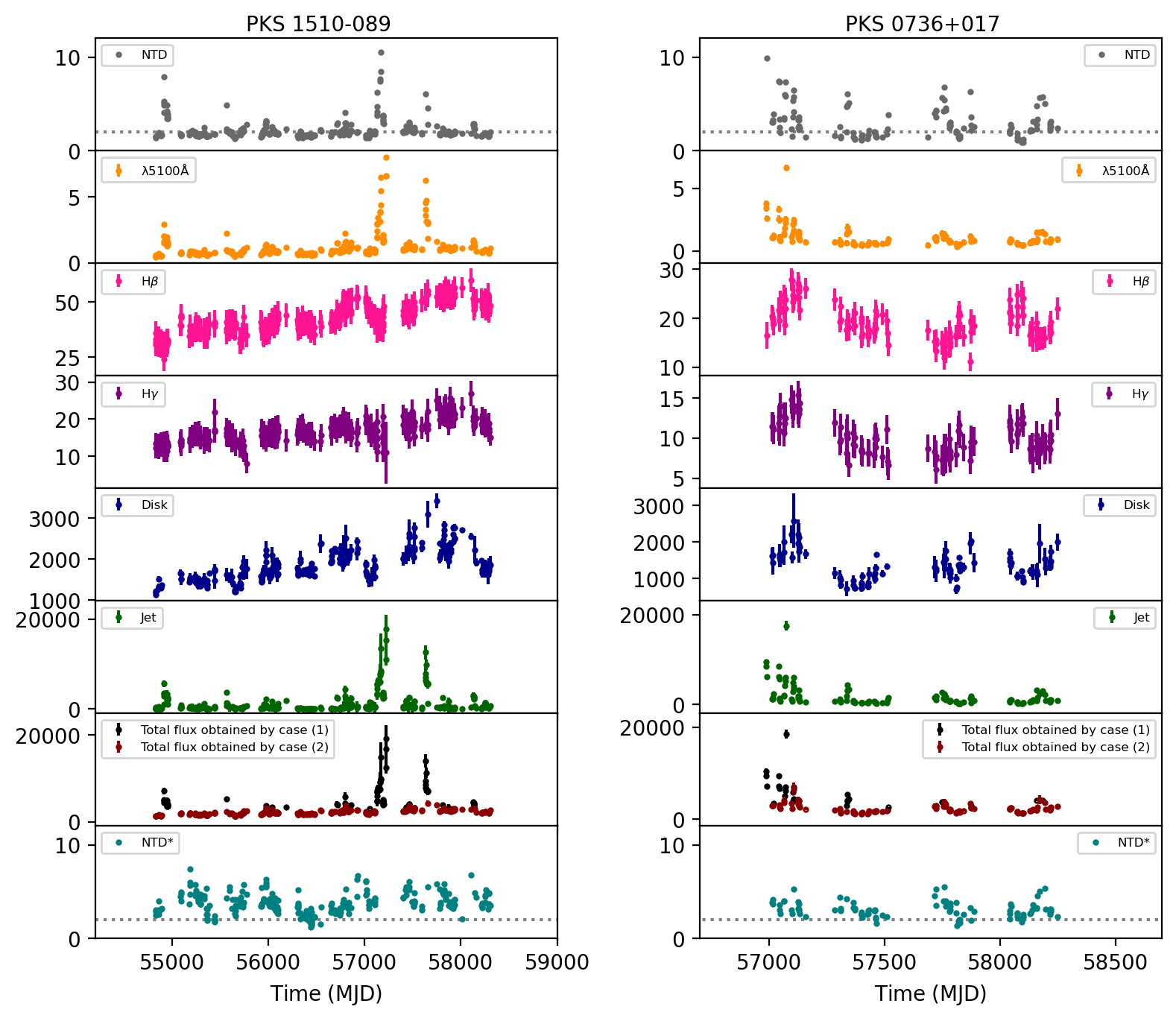}
    \caption{From top to bottom: light curves of the NTD parameter, $\lambda5100\text{\AA}$ continuum, broad emission lines (H$\beta$ and H$\gamma$), jet, accretion disk, total flux (disk+jet), and the NTD* parameter for PKS 0736+017 (right) and PKS 1510-089 (left). Grey dashed lines in the top (NTD) and bottom (NTD*) panels indicate NTD=2 and NTD*=2, respectively. The Y-axis units are as follows: NTD and NTD* are dimensionless, the $\lambda5100\,\text{\AA}$ continuum is in $10^{-15}\,\mathrm{erg\,cm^{-2}\,s^{-1}\,\AA^{-1}}$, all other panels are in $10^{-15}\,\mathrm{erg\,cm^{-2}\,s^{-1}}$. %The underlying data are provided in Tables \ref{tabpks1510} and \ref{tabpks0736}.
    }
    \label{fig:lc}
\end{figure*}

\begin{figure*}
	\centering
         \includegraphics[width=0.9\linewidth]{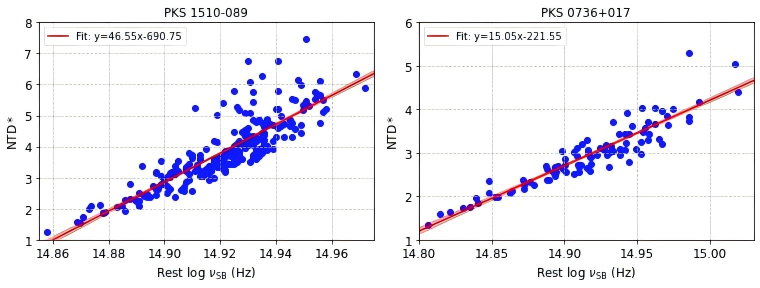}
    \caption{Relationship between the rest frequency of spectral break and the value of NTD*. The left panel shows PKS 1510-089, and the right panel shows PKS 0736+017. The red lines indicate the results of linear fits.} 
    \label{fig:fr}
\end{figure*}

In this section, we calculated the NTD parameter using both the conventional method and the results of the decomposition described in Section 2.3, and compared them.
Specifically, according to the conventional definition,
${\rm NTD}=L^{5100}_{\rm obs}/L^{5100}_{\rm disk}$, where $L^{5100}_{\rm obs}$ is the observed continuum luminosity at 5100 \AA, and $L^{5100}_{\rm disk}$ represents the disk luminosity at 5100\AA\; predicted from the H$\beta$ emission using the empirical relation provided by \citet{2020ApJS..249...17R}:
\begin{equation}
    {\rm log} L_{{\rm H}\beta}=(1.057\pm0.002){\rm log}L^{5100}_{\rm disk}+(-4.41\pm0.10).
\end{equation}
The variations of the NTD parameter are presented in the top panel of Figure \ref{fig:lc}, where the gray dashed line marks NTD$=$2.
On the other hand, as described in Section 2.3, we decomposed the disk and jet components by fitting the spectral break. This method was successfully applied to most observation IDs for the two blazars. Accordingly, we also calculated the NTD parameters based on the disk flux obtained from the spectral break fitting. To distinguish them from the conventional estimates, we denote the parameters derived from this method as NTD*, which is given by 
\begin{equation}
{\rm NTD^*}=(L_{\rm jet}^{5100}+L_{\rm disk}^{5100})/L_{\rm disk}^{5100}.
\end{equation}
The bottom panel in Figure \ref{fig:lc} shows the results, and the dashed line also marks NTD*$=$2. By comparing the two panels, it can be seen that for both blazars, very few observation IDs have NTD*$<$2, whereas nearly half of the observations yield NTD$>$2. This indicates that the disk flux estimated from Equation (1) may be overestimated in blazars. In addition, Figure \ref{fig:fr} shows the variation of the spectral break frequency $\nu_{\rm SB}$ with NTD* for the two blazars. It can be seen that when NTD*$\gtrsim$6 for PKS 1510-089 and NTD*$\gtrsim$5 for PKS 0736+017, the spectral break is difficult to be observed within the optical window. %At this stage, the corresponding luminosity ratio of the jet to the disk is approximately 4:1.}

\section{Reverberation Mapping} \label{sec:3}

%\subsection{Light curves} \label{sec:3.1}

%Based on the data analysis and processing described in Section \ref{sec:2}, we obtained the emission-line, jet, and accretion disk light curves for the two blazars as shown in Figure 4. To assess the reliability of the disk–jet decomposition presented in Section 2, we carried out the correlation analysis. A successful decomposition is expected to show that the emission-line light curves lag behind those of the accretion disk, consistent with the AGN unified model.

In this work, we first employed the ICCF method \citep{1998ApJ...501...82P} for the correlation analysis, following the procedure described in \citet{2022MNRAS.516.2671P}. Figure \ref{fig:iccf1510} shows the correlation between the disk and H$\beta$, as well as the correlation between the disk and H$\gamma$ for PKS 1510–089. Figure \ref{fig:iccf0736} presents the corresponding results for PKS 0736+017. In both sources, the variations of H$\beta$ and H$\gamma$ are found to lag behind those of the disk. Second, we applied the ROA method \citep{2021MNRAS.508.5449D} for comparison. Figures \ref{fig:PyROA1510} and \ref{fig:PyROA0736} show the ROA results for PKS 1510–089 and PKS 0736+017, respectively. The derived time lags for the both methods are summarized in Table \ref{tab:lag_results}. The results from the ICCF and ROA methods are consistent within uncertainties, with ROA providing slightly smaller errors.

%We computed the centroid lag ($\tau_c$) for each source. For PKS 1510-089, we obtained $\tau_c = -115.9^{+38.3}_{-32.1}$ days. For PKS 0736+017, the corresponding values are $\tau_c = -52.2^{+32.2}_{-21.8}$ days. These results are consistent with expectations that, in most AGNs, the variability of broad emission lines typically lags behind that of the accretion disk continuum by several tens of days. 

%\begin{figure*}
%	\centering
%        \includegraphics[width=0.45\linewidth]{Lag0736_CCF (1).png}
%         \includegraphics[width=0.45\linewidth]{Lag0736_P.png}
%        \includegraphics[width=0.45\linewidth]{Lag1510_CCF (1).png}
%         \includegraphics[width=0.45\linewidth]{Lag1510_P.png}
%    \caption{Top panels: The ICCF (left panel) and centroid probability distribution (right panel) between the accretion disk continuum and the H$\beta$ emission line in PKS 0736+017. Bottom panels: The ICCF (left panel) and centroid probability distribution (right panel) between the accretion disk continuum and the H$\gamma$ emission line in PKS 1510-089. The negative lag indicates that the former leads the latter.} 
%    \label{fig:ccf}
%\end{figure*}

\begin{table*}[]
\centering
\caption{Results of lag measurements (in days) for PKS 1510-089 and PKS 0736+017.}
\begin{tabular}{ccccc}
\hline
      & \multicolumn{2}{c}{PKS 1510-089}                   & \multicolumn{2}{c}{PKS 0736+017}             \\
      & Disk vs $\rm H\beta$     & Disk vs $\rm H\gamma$   & Disk vs $\rm H\beta$ & Disk vs $\rm H\gamma$ \\ \hline
ROA & $-110.9^{+16.2}_{-16.0}$ & $-94.1^{+13.5}_{-12.9}$ &     $-67.1^{+11.6}_{-11.9}$                 &     $-65.8^{+21.1}_{-19.0}$                  \\
ICCF  & $-134.3^{+36.2}_{-37.0}$  &  $-91.3^{+37.4}_{-43.1}$  &   $-53.2^{+23.3}_{-29.4}$                    &    $-59.9^{+21.5}_{-17.5}$                    \\ \hline
\end{tabular}
\label{tab:lag_results}
\end{table*}

\begin{figure*}
	\centering
        \includegraphics[width=0.99\linewidth]{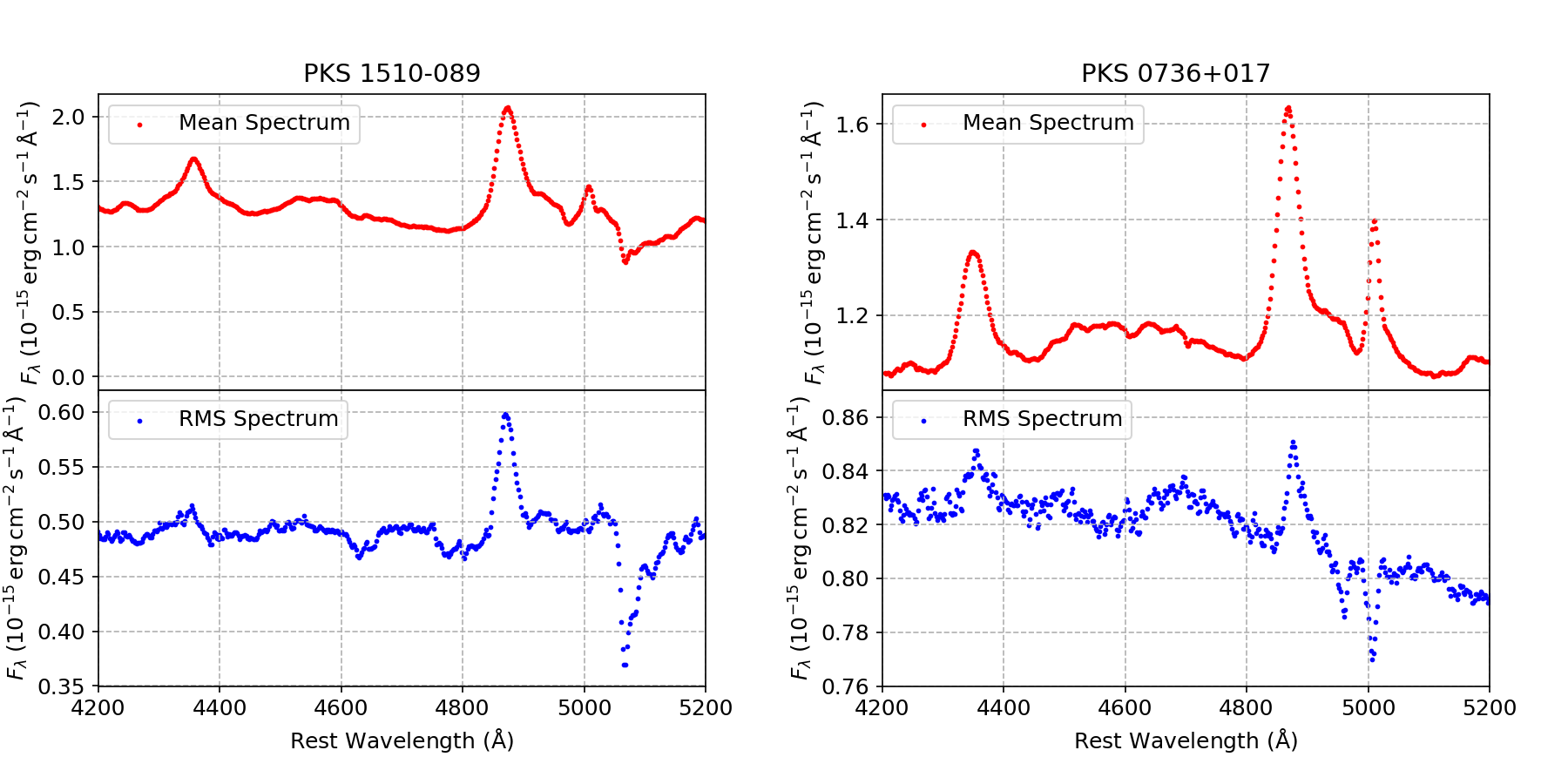}
    \caption{Mean (top panels, red dots) and RMS (bottom panels, blue dots) spectra for the blazars PKS 1510-089 (left) and PKS 0736+017 (right).} 
    \label{fig:fwhm}
\end{figure*}

To estimate the black hole masses, we measured the FWHM and line dispersion ($\sigma_{\rm line}$) of the emission lines from both the mean and root-mean-square (RMS) spectra, following the procedure described in \citet{2004ApJ...613..682P}. The mean and RMS spectra for the two blazars are shown in Figure \ref{fig:fwhm}. As evident in the RMS spectra, the H$\gamma$ line is very weak, rendering its measurements unreliable. Therefore, in the RMS spectra, we only measured the FWHM and $\sigma_{\rm line}$ for H$\beta$. In the mean spectra, we measured the FWHM and $\sigma_{\rm line}$ for both H$\gamma$ and H$\beta$. The measured values of both FWHM and $\sigma_{\rm line}$ for the two blazars are summarized in Table \ref{tab:bh_results}.

\begin{table}[]
\centering
\caption{Measurements of emission line widths and black hole masses obtained from the mean and rms spectra.}
\begin{tabular}{cc|ccc|ccc}
\hline
                                           &                            & \multicolumn{3}{c|}{PKS 1510-089} & \multicolumn{3}{c}{PKS 0736+017} \\
                                           &                            & Type                 & $\Delta V\;(\rm km\;s^{-1})$  & $M_{\rm BH}\;(10^{7}M_\odot$)  & Type       & $\Delta V\;(\rm km\;s^{-1})$      & $M_{\rm BH}\;(10^{7}M_\odot$)      \\ \hline
\multicolumn{1}{c}{\multirow{4}{*}{Mean}} & \multirow{2}{*}{H$\beta$}  & FWHM                 &  $2295\pm243$   &  $12.6^{+3.6}_{-3.0}$    & FWHM            & $1938\pm278$         &   $5.5^{+2.1}_{-1.6}$       \\
\multicolumn{1}{c}{}                      &                            & $\sigma_{\rm line}$  &  $1250\pm120$   &  $14.9^{+3.9}_{-3.3}$    &    $\sigma_{\rm line}$        &  $1139\pm121$        &  $7.5^{+2.3}_{-1.9}$         \\
\multicolumn{1}{c}{}                      & \multirow{2}{*}{H$\gamma$} & FWHM                 &  $2335\pm260$    &  $11.1^{+3.2}_{-2.7}$    &  FWHM          &  $2731\pm254$         &    $10.5^{+4.2}_{-3.6}$      \\
\multicolumn{1}{c}{}                      &                            & $\sigma_{\rm line}$  &  $1446\pm132$   &  $17.0^{+4.3}_{-3.7}$    &  $\sigma_{\rm line}$          &  $1379\pm139$        &   $10.7^{+4.4}_{-3.6}$        \\ \hline
\multicolumn{1}{c}{\multirow{4}{*}{RMS}}  & \multirow{2}{*}{H$\beta$}  & FWHM                 & $2073\pm121$    &  $10.4^{+2.0}_{-1.8}$    &    FWHM         &  $2054\pm189$        &  $6.2^{+1.5}_{-1.3}$        \\
\multicolumn{1}{c}{}                      &                            & $\sigma_{\rm line}$  & $1298\pm132$    &  $16.1^{+4.5}_{-3.7}$     &   $\sigma_{\rm line}$         &   $1207\pm138$       &    $8.4^{+2.7}_{-2.3}$        \\
\multicolumn{1}{c}{}                      & \multirow{2}{*}{H$\gamma$} & FWHM                 & -   &   -   &   FWHM       & -        & -        \\
\multicolumn{1}{c}{}                      &                            & $\sigma_{\rm line}$  &   -   &  -    & $\sigma_{\rm line}$          & -        & -        \\ \hline
\end{tabular}
\label{tab:bh_results}
\end{table}

Using the measured time lags together with the FWHM or $\sigma_{\rm line}$ values, we estimated the black hole masses of both blazars. The black hole mass is calculated using the formula from \citet{2004ApJ...613..682P}:
\begin{equation}
    M_{\rm BH} = f \frac{R_{\rm BLR} \Delta V^2}{G} = f \frac{c\tau\Delta V^2}{G},
\end{equation}
where $f$ is set to be 1.12 for FWHM and 4.47 for $\sigma_{\rm line}$ \citep{2015ApJ...801...38W}; $c$ is the speed of light; $R_{\rm BLR} = c \tau$ is the BLR size; $G$ is the gravitational constant; and $\Delta V$ represents the velocity width of the broad line. Since the time lag measured with ROA provides a smaller uncertainty, we adopted the $\tau$ values of ROA for the black hole mass calculation. The measured black hole masses based on different emission lines and using both FWHM and $\sigma_{\rm line}$ values are summarized in Table \ref{tab:bh_results}. For PKS 1510-089, the derived average black hole mass is $1.4\times10^{8}\,M_{\odot}$, while for PKS 0736+017, the average mass is $8.1\times10^{7}\,M_{\odot}$. Following \cite{2000ApJ...533..631K} and \cite{2002A&A...387..422B}, we estimated the mass accretion rates normalized by the Eddington accretion rate ($\dot{m}$) for both sources, using the black hole masses and the average disk luminosities at 5100 \AA. The resulting normalized accretion rate is $\dot{m} \approx 0.26$ for PKS 1510-089 and $\dot{m} \approx 0.14$ for PKS 0736+017.

%For PKS 1510-089, the mean FWHM of the H$\gamma$ line is 2693 km/s. For PKS 0736+017, the mean FWHM of the H$\beta$ line is 2470 km/s. Based on these values and the measured time lags, the estimated black hole masses are $1.8^{+0.5}_{-0.6} \times 10^8\ M_\odot$ for PKS 1510-089 and $7.0^{+2.9}_{-4.3} \times 10^7\ M_\odot$ for PKS 0736+017, respectively.

%\begin{figure*}
%	\centering
%        %\includegraphics[width=0.49\linewidth]{2.png}
%         \includegraphics[width=0.49\linewidth]{RL.png}
%        %\includegraphics[width=0.49\linewidth]{LCCF0736.png}
%         %\includegraphics[width=0.49\linewidth]{LCCF1510.png}
%    \caption{} 
%    \label{fig:rl}
%\end{figure*}

\section{Discussion and conclusion} \label{sec:4}
Previous studies have made efforts to perform reverberation mapping on the blazars PKS 1510-089 and PKS 0736+017, both of which relied on the NTD parameter to separate disk and jet contributions. For PKS 1510-089, \cite{2024ApJ...977..178A} performed a correlation analysis between the H$\beta$ line and the $5100\,\text{\AA}$ continuum after excluding data points with $\mathrm{NTD} > 2$, and reported a time lag of about 80 days. This approach is feasible primarily because that the accretion disk in PKS 1510–089 is intrinsically bright, with only about 39\% of the data points showing $\mathrm{NTD} > 2$. In contrast, for PKS 0736+017, \cite{2022MNRAS.516.2671P} also calculated the NTD values and found that the average $\mathrm{NTD}$ is as high as 2.7. This indicates that the majority of the $5100\,\text{\AA}$ data points are jet-dominated. If all data points with $\mathrm{NTD} > 2$ are excluded, more than half (about 67\%) of the dataset would be removed, making it difficult to conduct a reliable correlation analysis.
In our algorithm, the filtered data points with valid disk fluxes are 72\% of all measurements for PKS 0736+017. Thus, our algorithm excludes only 28\% of the data, significantly fewer than the 67\% of the data under the NTD-based selection. 
%Also, the peaks of ICCF for both targets exceed 0.6, which also produce significant results than the previous RM methods for blazars. 

%One valid condition of our method is that the targets show spectral break at the optical window. For blazars such as 3C 454.3 and 3C 279, where the optical–UV continuum is almost entirely jet-dominated and the thermal “big blue bump” component is virtually undetectable \citep{2021ApJS..257...37P}, our method does not work. The fraction of the valid targets for the blazar RM will be answered for a sample work in the future.

Previous studies have reported a wide range of black hole mass estimation for these two blazars based by other methods. For PKS 1510-089, by modeling the temperature profile of the accretion disk, \cite{2010ApJ...721.1425A} and \cite{2017A&A...601A..30C} inferred the black hole mass to be $5.4 \times 10^8\ M_\odot$ and $2.4 \times 10^8\ M_\odot$, respectively. In addition, using the single-epoch spectrum method, \cite{2002ApJ...576...81O} and \cite{2005AJ....130.2506X} derived masses of $3.9 \times 10^8\ M_\odot$ and $2.0 \times 10^8\ M_\odot$ from the $\lambda5100$ continuum luminosity and the FWHM of the H$\beta$ line. In comparison, the black hole mass of PKS 1510-089 from our RM ($1.4 \times 10^8\ M_\odot$) is relatively smaller than those derived by other methods. For PKS 0736+017, \cite{2001MNRAS.327..199M} and \cite{2002ApJ...579..530W} presented a black hole mass of $2.9 \times 10^8\ M_\odot$ and $1.0 \times 10^8\ M_\odot$ by using bolometric luminosity. Our RM measurement yields $8.1 \times 10^7\ M_\odot$, which is also relatively smaller. This discrepancy may be due to an overestimation of the disk luminosity in blazars by those methods.

In summary, we proposed a novel approach to perform the reverberation mapping method for blazars by disentangling the disk and jet components through modeling the spectral break  in the log $\nu$ versus log $\nu F_\nu$ plane. Applying this method to PKS 1510-089 and PKS 0736+017, we obtained the light curves of the accretion disk and emission lines, and obtained the fiducial lags between them. In PKS 1510-089, the H$\gamma$ variability lags behind the accretion disk by approximately 94 days, while the H$\beta$ line shows a lag of about 111 days relative to the disk. In PKS 0736+017, the H$\gamma$ variability lags behind the disk by roughly 66 days, and the H$\beta$ line exhibits a lag of about 67 days. Based on these measured time lags, we estimate black hole masses of $\sim1.4\times10^{8}\,M_{\odot}$ for PKS 1510-089 and $\sim8.1\times10^{7}\,M_{\odot}$ for PKS 0736+017, respectively. Compared to traditional NTD-based selection methods, our approach retains more data and provides more significant lag estimation. Thus, our method paves a way to measure the black hole mass for specific jet-dominated AGNs.

\begin{acknowledgments}
The data underlying this article were accessed from the Steward Observatory data base with the following link: http://james.as.arizona.edu/$\sim$psmith/Fermi/ (see, \citealt{2009arXiv0912.3621S}). This work has been funded by the National Natural Science
Foundation of China under grant No. U2031102, and the Shandong Provincial Natural Science Foundation under grant No. ZR2023MA036.
\end{acknowledgments}

\software{agnpy \citep{2022A&A...660A..18N}, SciPy \citep{2020NatMe..17..261V},PyROA \citep{2021MNRAS.508.5449D}}

%% Appendix material should be preceded with a single \appendix command.
%% There should be a \section command for each appendix. Mark appendix
%% subsections with the same markup you use in the main body of the paper.
%%
%% Each Appendix (indicated with \section) will be lettered A, B, C, etc.
%% The equation counter will reset when it encounters the \appendix
%% command and will number appendix equations (A1), (A2), etc. The
%% Figure and Table counter will not reset.

\appendix
\counterwithin{figure}{section}
\counterwithin{table}{section}
\renewcommand{\thefigure}{A\arabic{figure}}
\renewcommand{\thetable}{A\arabic{table}}

\section{Multi-wavelength SED model}
We employed the traditional one-zone leptonic model to reproduce the broadband SEDs presented in bottom panels of Figure \ref{fig:guangpuduoshiqi}. This model, widely used for blazars, assumes a homogeneous spherical emission region of radius \( R \), embedded in a magnetic field \( B \). The region moves relativistically along the jet  with a velocity \( v = \beta c \), corresponding to a bulk Lorentz factor \( \Gamma = (1 - \beta^2)^{-1/2} \). A Doppler factor \( \delta = [\Gamma(1 - \beta \cos\theta)]^{-1} \approx \Gamma \) is adopted, assuming a small viewing angle \( \theta < 1/\Gamma \), typical for blazar jets. The radiative processes include the synchrotron emission, the synchrotron self-Compton (SSC), and external Compton (EC) process. The calculation of spectrum refers to \citet{1979rpa..book.....R}.

Following \cite{2016MNRAS.461.1862K,2022MNRAS.513..611K,2023MNRAS.521.6210D}, we assumed that the electron energy distribution in the emission zone follows a broken power-law, characterized by spectral indices $p_1$ and $p_2$ below and above the broken Lorentz factor $\gamma_b$, respectively. The distribution is expressed as:
\begin{equation}
N(\gamma)=
\begin{cases}
N_0\gamma^{-p_1} & \gamma_{\rm min}\leq\gamma\leq\gamma_{b},\\
N_0\gamma_b^{p_2-p_1}\gamma^{-p_2} & \gamma_b<\gamma\leq\gamma_{\rm max}
\end{cases}    
\end{equation}
where $\gamma_{\rm min}$ and $\gamma_{\rm max}$ are the minimal and maximal electron Lorentz factors, and $N_0$ is the normalized particle number density.  

For the EC emission, the seed photons are assumed to originate from both the BLR and the DT. Following \cite{2022MNRAS.513..611K,2024PASP..136l4101D}, we model the photon fields from the BLR and DT as single-temperature blackbody spectra with temperatures of $T_{\rm BLR} = 4.2 \times 10^4$\,K and $T_{\rm DT} = 10^3$\,K, respectively. Their photon energy densities, $u_{\rm BLR}$ and $u_{\rm DT}$, are treated as free parameters.

We model the accretion disk as a geometrically thin, optically thick Shakura--Sunyaev disk \citep{1973A&A....24..337S}, using the \text{agnpy}\footnote{\url{https://agnpy.readthedocs.io/en/latest/index.html}} Python package \citep{2022A&A...660A..18N}. The inner and outer disk radii are set to the default values, \( R_{\rm in} = 6\,R_{\rm g} \) and \( R_{\rm out} = 200\,R_{\rm g} \), where \( R_{\rm g}\) is the gravitational radius. The radiative efficiency is also fixed \( \eta = 1/12 \), representing the fraction of accreted mass energy converted into radiation. As a result, the accretion disk component only has two free parameters: the black hole mass ($M_{\rm BH}$) and the disk luminosity ($L_{\rm disk}$).

To reproduce the broadband SED, a total of thirteen parameters are required: \( N_0 \), \( \gamma_{\rm min} \), \( \gamma_b \), \( \gamma_{\rm max} \), \( p_1 \), \( p_2 \), \( B \), \( R \), \( \delta \), \( u_{\rm DT} \), \( u_{\rm BLR} \), \( L_{\rm disk} \), and \( M_{\rm BH} \).  
Given the complexity of the parameter space, applying rigorous optimization methods such as MCMC would be computationally expensive and difficult to converge. Thus, the model parameters were tested within feasible values to fit the SEDs 
by eye estimation such as \citet{2021MNRAS.504.1103R}.
Finally, for PKS 1510-089, the adopted model parameters are as follows:  
\( N_0 = 398\;\rm cm^{-3}\),  
\( \gamma_{\rm min} = 300 \),  
\( \gamma_{b} = 500 \),  
\( \gamma_{\rm max} = 10000 \),  
\( p_1 = 1.3 \),  
\( p_2 = 2.3 \),  
\( B = 0.016\;\rm G\),  
\( R = 0.01\;{\rm pc}\),  
\( \delta = 32 \),  
\( u_{\rm DT} = 9\times10^{-5}\;\rm erg\;cm^{-3}\),  
\( u_{\rm BLR} = 3\times10^{-5}\;\rm erg\;cm^{-3}\),  
\( L_{\rm disk} = 4\times10^{45}\;\rm erg\;s^{-1} \),  
\( M_{\rm BH} =3\times10^{8}\;M_\odot\).
For PKS 0736+017, the adopted model parameters are as follows:  
\( N_0 = 251\;\rm cm^{-3}\),  
\( \gamma_{\rm min} = 10 \),  
\( \gamma_{b} = 300 \),  
\( \gamma_{\rm max} = 4600 \),  
\( p_1 = 1.3 \),  
\( p_2 = 2.3 \),  
\( B = 0.25\;\rm G\),  
\( R = 0.01\;{\rm pc}\),  
\( \delta = 11 \),  
\( u_{\rm DT} = 4\times10^{-3}\;\rm erg\;cm^{-3}\),  
\( u_{\rm BLR} = 3\times10^{-3}\;\rm erg\;cm^{-3}\),  
\( L_{\rm disk} = 4\times10^{45}\;\rm erg\;s^{-1} \),  
\( M_{\rm BH} =2\times10^{8}\;M_\odot\).

\section{Plots of ICCF and ROA for PKS 1510-089 and PKS 0736+017}

\begin{figure*}
	\centering
        \includegraphics[width=0.9\linewidth]{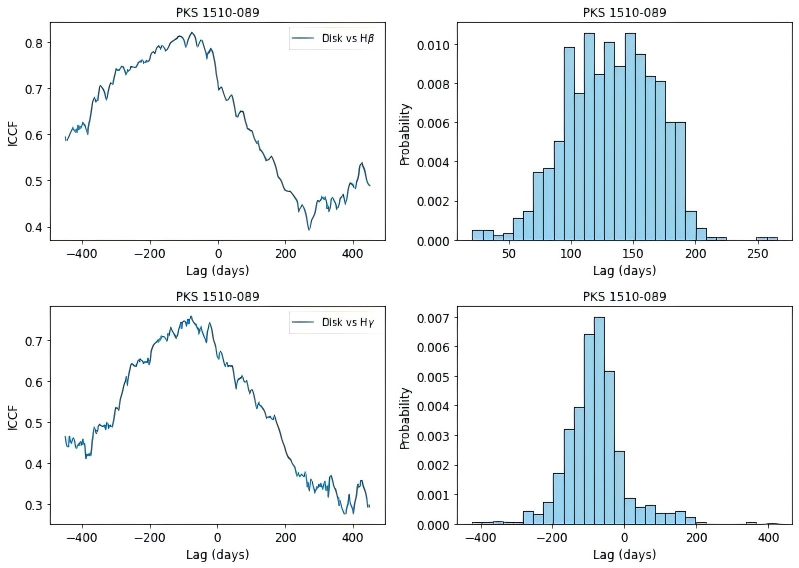}
    \caption{Top panels: ICCF (left) and centroid probability distribution (right) between the disk continuum and the H$\beta$ line in PKS 1510-089.
Bottom panels: Same as above, but for the H$\gamma$ line.} 
    \label{fig:iccf1510}
\end{figure*}

\begin{figure*}
	\centering
        \includegraphics[width=0.9\linewidth]{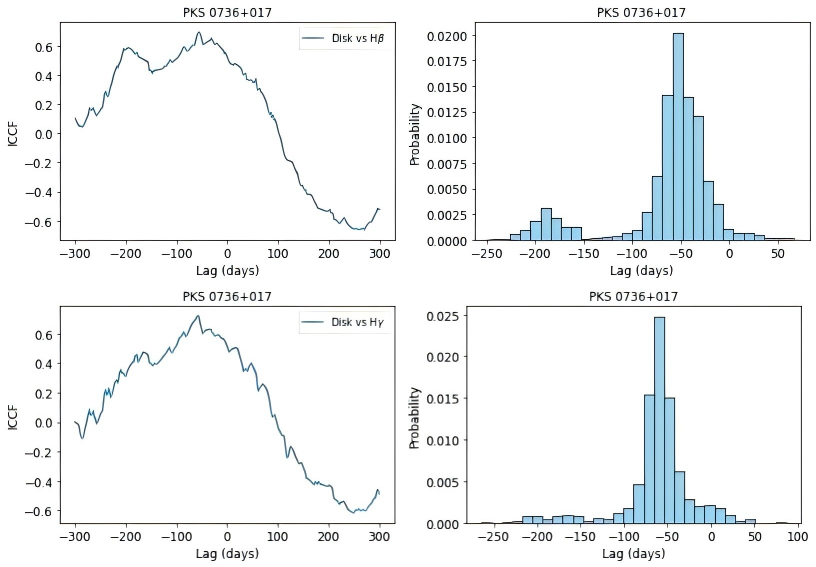}
    \caption{Top panels: ICCF (left) and centroid probability distribution (right) between the disk continuum and the H$\beta$ line in PKS 0736+017.
Bottom panels: Same as above, but for the H$\gamma$ line.} 
    \label{fig:iccf0736}
\end{figure*}

\begin{figure*}
	\centering
        \includegraphics[width=0.99\linewidth]{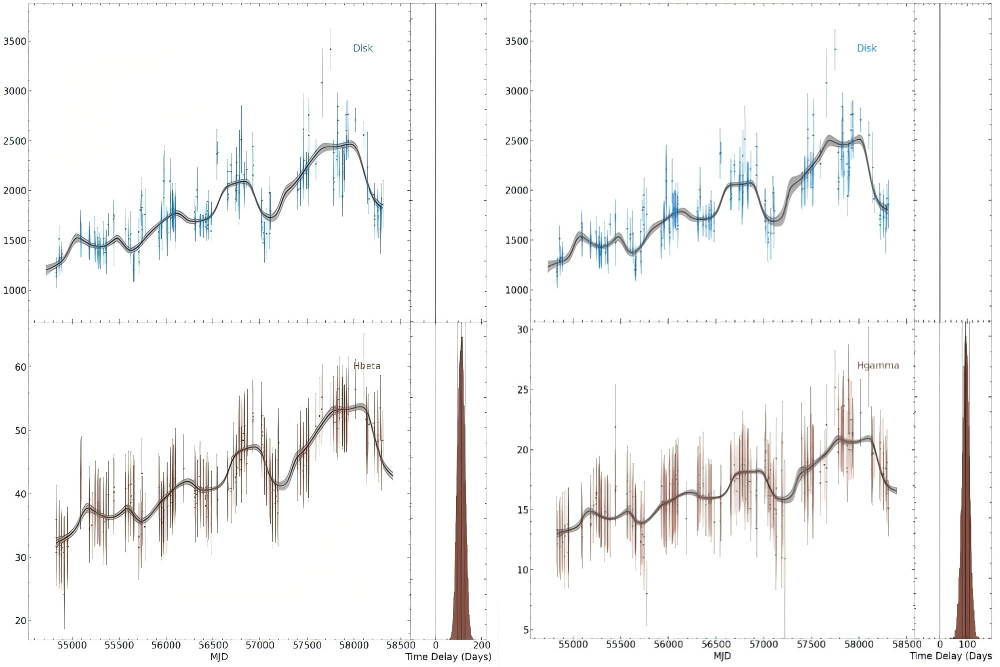}
    \caption{The figure shows the PyROA output for PKS 1510-089. The left panel displays the result between the disk and the H$\beta$ line, while the right panel shows the result with the H$\gamma$ line.} 
    \label{fig:PyROA1510}
\end{figure*}

\begin{figure*}
	\centering
        \includegraphics[width=0.99\linewidth]{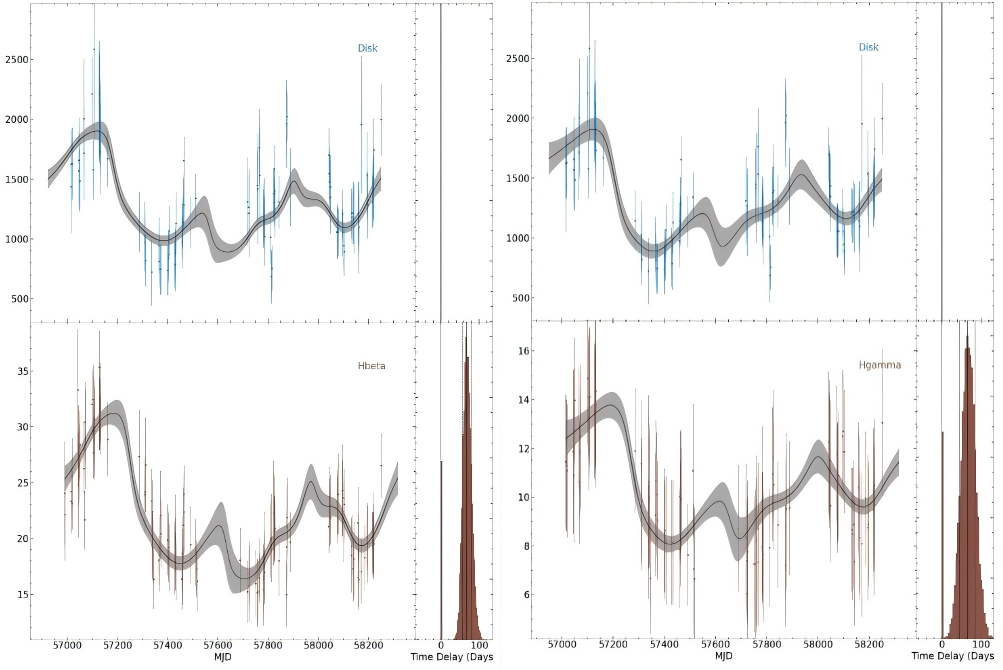}
    \caption{The figure shows the PyROA output for PKS 0736+017. The left panel displays the result between the disk and the H$\beta$ line, while the right panel shows the result with the H$\gamma$ line.} 
    \label{fig:PyROA0736}
\end{figure*}

%LCCF involves progressively shifting the light curve of one component relative to another and computing the cross-correlation at each step. The LCCF is mathematically defined as follows:
%\begin{equation}
%    \text{LCCF}(\tau) = \frac{1}{M} \sum_{ij} \frac{(a_i - \bar{a}_{\tau})(b_j - \bar{b}_{\tau})}{\sigma_{a\tau} \sigma_{b\tau}},
%\end{equation}
%where \(a_i\) and \(b_j\) represent time series, and \(M\) denotes the number of \((a_i, b_j)\) pairs that satisfy the condition \(\tau \leq \Delta t_{ij} \leq \tau + \delta t\) (where \(\delta t\) is the bin time). The averages of the \(M\) pair samples are denoted by \(\bar{a}_{\tau}\) and \(\bar{b}_{\tau}\), while \(\sigma_{a\tau}\) and \(\sigma_{b\tau}\) represent the corresponding standard deviations. The uncertainties in the LCCF coefficient are considered to be the standard deviation of the local \(M\) samples, i.e.,
%\begin{equation}
%%    \sigma_{\text{LCCF}(\tau)} = \left( \frac{1}{M-1} \sum \left[ \text{LCCF}_{ij} - \text{LCCF}(\tau) \right]^2 \right)^{1/2}.
%\end{equation}
%The values of the LCCF range from \([-1, 1]\), which is a desirable property for significance estimation.

\bibliography{sample7}{}
\bibliographystyle{aasjournalv7}

%% This command is needed to show the entire author+affiliation list when
%% the collaboration and author truncation commands are used.  It has to
%% go at the end of the manuscript.
%\allauthors

%% Include this line if you are using the \added, \replaced, \deleted
%% commands to see a summary list of all changes at the end of the article.
%\listofchanges

\end{document}